\documentclass[a4paper]{panl}
\usepackage{cite}
\usepackage{wrapfig}
\usepackage{graphicx}
\usepackage{amssymb}
\usepackage{amsfonts}
\usepackage{amsmath}
\usepackage{bm}
\usepackage{longtable}
\usepackage{rotating}
\usepackage{lscape}
\usepackage{epsfig}
\usepackage{multirow}

\usepackage{xcolor}
\usepackage{comment}

\newcommand{\mk}{ \bm k }
\newcommand{\mv}{ \bm v }
\newcommand{\mx}{ \bm x }

\newcommand{\mF}{ \bm F }
\def\eRM{\mathrm{e}}

\originalTeX
\begin{document}

\title{Anomalous kinetics of a multi-species reaction-diffusion system: effect of random velocity fluctuations}

\maketitle
\authors{M.\,Hnati\v{c} $^{a,b,c,} $\footnote{E-mail: hnatic@saske.sk},
M.\,Kecer $^{a,}$\footnote{E-mail: matej.kecer@student.upjs.sk}, T.\,Lu\v{c}ivjansk\'{y} $^{a,}$\footnote{E-mail: tomas.lucivjansky@upjs.sk}}
\setcounter{footnote}{0}
\from{$^{a}$\, Institute of Physics, Faculty of Science, P. J. \v{S}af\'{a}rik University, Park Angelinum 9, 040 01 Ko\v{s}ice, Slovakia}
\from{$^{b}$\, Institute of Experimental Physics, Slovak Academy of Sciences, Watsonova 47, 
040 01 Ko\v{s}ice, Slovakia}
\from{$^{c}$\, Joint Institute for Nuclear Research, 141980 Dubna, Russia}


\begin{abstract}

\vspace{0.2cm}
Reaction-diffusion systems, which consist of the reacting particles subject to diffusion process, constitute one of the common examples of non-linear statistical systems. In low space dimensions $d \leq 2$ the usual description by means of kinetic rate equations is not sufficient and the effect of density fluctuations has to be properly taken into account. Our aim here is to analyze a particular multi-species reaction-diffusion system characterized by reactions $\textit{A} +\textit{A} \rightarrow (\emptyset, A),$ $\textit{A} +\textit{B} \rightarrow \textit{A}$ at and below its critical dimension $d_c = 2$. In particular, we investigate effect of thermal fluctuations on the reaction kinetics, which are generated by means of random velocity field modelled by a stochastic Navier-Stokes equations. Main theoretical tool employed is  field-theoretic perturbative renormalization group. 
The analysis is performed to the first order of the perturbation scheme 
(one-loop approximation).
\end{abstract}
\vspace*{6pt}

\label{sec:intro}
\section*{Introduction and description of the model}

Investigation of the reaction kinetics has been attracting a lot of attention in the past 20 years \cite{tauber:2014}. 
One of the paradigmatic models is given by
a two-species reaction-diffusion system with reactions 
$\textit{A} +\textit{A} \rightarrow (\emptyset, A),$ $\textit{A} +\textit{B} \rightarrow \textit{A}$, in which the coagulation takes place with probability $p$ and annihilation with probability $1-p$. Although various aspects and variants of the model were already published, e.g. \cite{rajesh:2004, lee:2017, hellerick:2020}, the influence of velocity-field induced fluctuations on its kinetics remains yet unknown. As such effects are naturally present in many chemical and physical systems their investigation constitutes a major motivation for the present work.

For brevity, let us mention that given system can be conveniently
 recast in a form of field-theoretic action using the so-called Doi-Peliti formalism \cite{doi:1976, peliti:1985}.
The resulting actions takes the form
\begin{align}
     S_{\psi} &= \psi^\dagger_A(-\partial_t+\nu_0 u_{A0} \partial^2) \psi_A^{~} +  \psi^\dagger_B(-\partial_t+\nu_0 u_{B0}\partial^2) \psi_B^{~} - \nu_0 u_{A0} \lambda_0 \psi_A^\dagger \psi_A^2 \nonumber\\ 
     &- \nu_0 u_{A0} \lambda_0 \psi_A^{\dagger 2}\psi_A^2 - \lambda_0' Q \nu_0 u_{A0} \psi_B^\dagger \psi_A^{~} \psi_B^{~} - \nu_0 u_{A0} \lambda_0'\psi_A^\dagger \psi_B^\dagger \psi_A^{~} \psi_B^{~}, \label{S_psi}\\
     &+(\psi_A^\dagger \  a_0 + \psi_B^\dagger \ b_0) \delta(t), \nonumber
\end{align}
where $\partial^2$ denotes Laplace operator in $d$-dimensions, diffusion parameters are expressed through Prandtl numbers $u_{A0}$ and $u_{B0}$ and viscosity $\nu_0$, respectively.
The parameters $\lambda_0, \lambda'_0$ denote reaction constants and parameter $Q = 1/(2-p)$ is related to probability of whether annihilation or coagulation process takes place.
Last two terms in the action \eqref{S_psi} correspond to initial conditions, where $a_0$ ($b_0$) describe initial number density of particle 
type $A$ $(B)$.
Throughout the paper, in expressions for action functional integrations over space and time variables are implied. The subscript "$0$" is used to denote bare parameters (as opposed to the renormalized parameters which we write without subscript).
In the present work, we mainly focus on a special case of equal diffusivities for both particle types, i.e. we consider $u_{A0} = u_{B0} = u_0$.
The more general case is deferred to later work.

In order to model the advection of this reaction-diffusion system by random fluid environment, we introduce a velocity field $\mv(\mx,t)$.
We assume that it is a random variable with zero mean satisfying the stochastic Navier-Stokes equation \cite{frisch:1995}
\begin{equation}
    \partial_t v_i + (v_j \partial_j) v_i = \nu_0 \partial^2 v_i - \partial_i P + F_i. \label{stoch_NS}
\end{equation}
Here, the summation over repeated indices is implied, $P$ denotes pressure, and $\mF$
corresponds to an external random force.
We further assume the force $\mF$ is a Gaussian white noise with zero mean and prescribed correlation function in the form
%
%
\begin{align}
    \langle F_i(t,\mx) F_j(t', \mx') \rangle &= \delta(t-t') \int \frac{d^dk}{(2\pi)^d} D_0 k^2 P_{ij}(\mk) \eRM^{i\mk \cdot (\mx-\mx')}. \label{force_correlator}
\end{align}
%
%
In the present work we consider the case of an incompressible fluid, hence the appearance of transverse projection operator $P_{ij}(\mk) = \delta_{ij} - k_i k_j /k^2$.
The force term described in \eqref{force_correlator} was originally studied in \cite{forster:1977} and it serves to generate fluctuations of the velocity field near thermal equilibrium. 
From practical point of view, assumed form of velocity statistics is both simple and instructive, since all the non-linearities present in the model become simultaneously logarithmic  in critical space dimension $d = d_c$.

The stochastic problem \eqref{stoch_NS}, \eqref{force_correlator} is equivalent to the field-theoretic model with the doubled set of fields
$\Phi = \{ \mv, \mv' \}$
described by the De Dominicis-Janssen action functional \cite{vasiliev:2004}
\begin{align}
      S_{v} = \frac{1}{2} D_0 \partial_i v_j'  \partial_i v_j' + v_i' \bigl( -\partial_t v_i - v_j \partial_j v_i + \nu_0 \partial^2 v_i \bigr). \label{S_v}
\end{align}
Actions \eqref{S_psi} and \eqref{S_v} need to be supplemented with interactions terms that couple together the scalar and velocity fields. 
Such effect might be conveniently achieved by the following replacement
$\partial_t \rightarrow \partial_t + v_i \partial_i$
in the action \eqref{S_psi}, that effectively corresponds to advective processes for reacting particles.
The full action functional of the theory then becomes
\begin{equation}
    S = S_{\psi} + S_{v} -\psi_A^\dagger v_i \partial_i \psi_A -\psi_B^\dagger v_i \partial_i \psi_B. \label{S_t}
\end{equation}

Perturbation theory of the model is constructed using the Feynman diagrammatic technique \cite{vasiliev:2004}. The bare propagators of the theory are determined by quadratic part of full action, and in frequency-momentum representation they are 
$ \langle \psi_A^{~} \psi_A^\dagger \rangle_0 = 1/(-i\omega + \nu_0 u_{0} k^2),$
$\langle v_i v_j\rangle_0 = D_0 k^2 P_{ij}(\mk)/(\omega^2+\nu_0^2 k^4),$
$ \langle \psi_B^{~} \psi_B^\dagger\rangle_0 = 1/(-i\omega + \nu_0 u_{0} k^2),$
$\langle v_iv'_j\rangle_0  = P_{ij}(\mk)/(-i\omega + \nu_0k^2).$  
%
%
The nonlinear part of the action defines interaction vertices with vertex factors
\begin{align}
    \begin{split}
        & V_{ \psi_A^\dagger \psi_A^{~} \psi_A^{~} } = -2\lambda_0 \nu_0 u_{0},\\
        & V_{ \psi_B^\dagger \psi_B^{~} \psi_A^{~} } = -\lambda'_0 \nu_0 u_{0} Q,\\
        & V_{ \psi_A^\dagger(\mk) \psi_A^{~} v_j  } = V_{ \psi_B^\dagger(\mk) \psi_B^{~} v_j}  = i k_j, \label{bare_vert}
    \end{split}
    \quad
    \begin{split}
        & V_{ \psi_A^\dagger \psi_A^\dagger \psi_A^{~} \psi_A^{~} }= - 4\lambda_0 \nu_0 u_{0},\\
        & V_{ \psi_A^\dagger \psi_B^\dagger \psi_A^{~} \psi_B^{~} } = -\lambda'_0 \nu_0 u_{0},\\
       &  V_{ v'_i(\mk) v_l v_j }  = i(k_l \delta_{ij} + k_j \delta_{il})/2.
    \end{split}
\end{align}
%
%
{\label{sec:renormalization}
\section*{Renormalization of the model}}

As perturbation theory displays UV divergences, they have to be properly analyzed and this can be done by RG method \cite{vasiliev:2004}.
The starting point is the analysis of the canonical dimensions of all fields and parameters.
To each quantity $F$ we assign its frequency dimension $d_F^\omega$, momentum dimension $d_F^k$ and the total canonical dimension $d_F = d_F^k + 2d_F^\omega$. All relevant dimensions for the action
\eqref{S_t} are listed in Tab.~\ref{tab:canonical_dimensions}.
For an arbitrary 1-particle irreducible Green's function (1PI), the total canonical dimension can be written as $d_\Gamma = d+2 - \sum_\Psi N_\Psi d_\Psi$, 
where the sum runs through all the types of fields $\Psi$, $N_\Psi$ denotes number of times given field appears in the particular 1PI function and $d_\Psi$ is a canonical dimension of the given field \cite{vasiliev:2004}. The UV divergences which require further treatment are those irreducible functions which have non-negative formal index of divergence $\delta_\Gamma = d_\Gamma|_{\epsilon=0}$, where we defined $\epsilon = 2-d$.
It is possible to show that the only divergent structures are those already present in the bare action, which implies that the model is multiplicatively renormalizable.
\begin{table*}
  \begin{center}
    \begin{tabular}{|c|c|c|c|c|c|c|c|c|c|}
      \hline
      $F$ & $\psi_A$, $\psi_B$ & $\psi^\dagger_A$, $\psi^\dagger_B$ & $\lambda_0$, $\lambda'_0$ & $u_{0}$, $Q$ & $a_0$, $b_0$ & $v$ & $v'$ & $\nu_0$ & $D_0$ \\
      \hline
      $d_F^k$ & $d$ & $0$ & $2-d$ & $0$ & $d$ & $-1$ & $d+1$ & $-2$ & $-d-4$ \\
      \hline
      $d_F^\omega$ & $0$ & $0$ & $0$ & $0$ & $0$ & $1$ & $-1$ & $1$ & $3$ \\
      \hline
      $d_F$ & $d$ & $0$ & $2-d$ & $0$ & $d$ & $1$ & $d-1$ & $0$ & $2-d$\\
      \hline
    \end{tabular}
  \end{center}
  \caption{Canonical dimensions of fields and parameters.}
  \label{tab:canonical_dimensions}
\end{table*}

Renormalized action can be written in the form
%
%
\begin{align}
    S_{R} &=  \psi^\dagger_A(-\partial_t+ Z_1 u \nu \partial^2) \psi_A^{~} +  \psi^\dagger_B(-\partial_t + Z_1 u \nu \partial^2) \psi_B^{~} + \frac{1}{2} Z_2 \mu^\epsilon D \partial_i v_j' 
 \partial_i v_j' \nonumber \\
    &+ v_i' (-\partial_t + Z_3 \nu \partial^2) v_i - Z_4 u \nu \mu^\epsilon \lambda \bigl[ \psi_A^\dagger \psi_A^2 + \psi_A^{\dagger 2}\psi_A^2 \bigr]\ \nonumber \\
    &- Z_5 u \nu \mu^\epsilon \lambda' \bigl[Q  \psi_B^\dagger \psi_A^{~} \psi_B^{~} + \psi_A^\dagger \psi_B^\dagger \psi_A^{~} \psi_B^{~} \bigr] 
    - v_i' ( \mv \cdot \boldsymbol{\partial} ) v_i \label{renorm_action} \\ 
    &- \bigl[ \psi_A^\dagger ( \mv \cdot \boldsymbol{\partial} ) \psi_A^{~}  + \psi_B^\dagger ( \mv \cdot \boldsymbol{\partial} ) \psi_B^{~} \bigr] +  \psi_A^\dagger \  a_0 + \psi_B^\dagger \ b_0 \nonumber.
\end{align}
%
%
Renormalization constants $Z_1 - Z_5$ appearing in \eqref{renorm_action} are calculated from divergent parts of one-loop Feynman diagrams using dimensional regularisation and minimal subtraction scheme.
They read
$ Z_1 = 1 - \hat{g}/4u(u+1)\epsilon$,\, 
$Z_4 = 1 + \hat{\lambda}/\epsilon,$\,
  $Z_2 = Z_3 = 1 - \hat{g}/16\epsilon,$\, 
  $Z_5 = 1 + \hat{\lambda}'/2\epsilon,$
where $\hat{F} = F S_d/(2\pi)^d$, $g$ is renormalized version of parameter defined as $g_0 = D_0/\nu_0^3$, and $S_d$ is the area of unit $d$-dimensional sphere.
The RG equation for renormalized Green functions $G$ is written as
$D_{RG} \ G(e, \mu, ...) = \bigl[ \mu \partial_\mu + \sum_e \beta_e \partial_e - \gamma_\nu \nu \partial_\nu \bigr]  G(e, \mu, ...)=0,$
where the sum runs through all charges of the theory $e = \{ g, u, \lambda', \lambda \}$ and coefficient functions are defined as
$\beta_e = \mu \partial_\mu e|_{ _0}$, $\gamma_F = \mu \partial_\mu \ln Z_F|_{_0}$,
for any parameter $F$.
For our model, the $\beta$-functions are 
 $ \beta_x = -g(\epsilon + \gamma_x), \beta_{u} = -u\gamma_{u},$
where $x\in \{g,\lambda,\lambda'\}$
with corresponding anomalous dimensions
\begin{align}
\gamma_g =& -\frac{\hat{g}}{8}, \quad
\gamma_{u} = \hat{g} \biggl( \frac{1}{4 u (1+u)} - \frac{1}{16}\biggr), \quad
\gamma_{\nu} = \frac{\hat{g}}{16},\nonumber \\
\gamma_\lambda =& -\hat{\lambda}- \frac{\hat{g}}{4u(1+u)},\quad
\gamma_{\lambda'} = - \frac{\hat{\lambda'}}{2} - \frac{\hat{g}}{4u(1+u)}.
\label{gamma_functions}
\end{align}

Fixed points (FP) of the RG equation are such points in the space of coupling constants for which all $\beta$-functions identically vanish.
Long-time asymptotic behavior of our model is governed
by IR stable FPs for which all eigenvalues of matrix $\Omega_{ij} = \partial \beta_i/\partial g_j$, have positive real parts.

We found eight FPs. However, only two are IR stable
%
%
\begin{enumerate}
    \item Gaussian fixed point: $\hat{g}^* = 0$, $u^* =$ arbitrary, $\hat{\lambda}^* = 0$, $\hat{\lambda}'^* = 0$. IR stable for $\epsilon<0$.
    \item Thermal fixed point: $\hat{g}^* = 8\epsilon$, $u^* =(-1+\sqrt{17})/2$, $\hat{\lambda}^* = \epsilon/2$, $\hat{\lambda}'^* = \epsilon$. IR stable for $\epsilon>0$.
\end{enumerate}
%
%
Notice that for non-trivial (thermal) FP both velocity fluctuations and reaction interactions are simultaneously relevant. 

There is also a FP for which only reactions are relevant, and although it would have been stable without the of velocity field, it can never be IR stable in a presence of advective processes.

\label{sec:conclusions}
\section*{Conclusions}
In this paper, we have investigated effects of thermal fluctuations on a 
 specific reaction-diffusion system. We have concentrated on a special limit, 
 in which both particle types diffuse with the same diffusion constant. RG analysis reveals existence of the two IR stable FPs, which are main
 candidates for macroscopically observed regimes. These are: Gaussian fixed point in space dimensions $d>2$ and thermal fixed point for $d<2$, for which both velocity fluctuations and reactions are simultaneously important.
We expect that on the borderline between these regimes ($d=2$), particle densities  experience logarithmic corrections. Also, it is permissible that higher-loop corrections
might change corresponding regions of stability. Explicit calculation of time-decay exponent of particle densities in stable regimes and on the borderline, as well as analysis of the more general case of unequal difusivities is still a work in progress.

\label{sec:acknowledgments}
\section*{Acknowledgments}
The work was supported by VEGA grant No. 1/0535/21 of the Ministry of Education, Science, Research and Sport of the Slovak Republic.

\end{document}